\documentclass[prb,aps,floats,twocolumn]{revtex4b3}
\usepackage{epsfig,amssymb,amsmath}
\begin{document}
%------------------------------------------------------------------
\title{Lineshape predictions via Bethe ansatz for the one-dimensional spin-1/2
  Heisenberg antiferromagnet in a magnetic field} 
\author{Michael Karbach$^*$ and Gerhard M{\"u}ller$^\dagger$} 
\affiliation{$^*$Bergische Universit{\"a}t Wuppertal, Fachbereich Physik,
  D-42097 Wuppertal, Germany \\
  $^\dagger$Department of Physics, University of Rhode Island, Kingston RI 02881-0817}
\date{\today~--~1.6}
\begin{abstract}
  The spin fluctuations parallel to the external magnetic field in the ground
  state of the one-dimensional (1D) $s=\frac{1}{2}$ Heisenberg antiferromagnet
  are dominated by a two-parameter set of collective excitations. In a cyclic
  chain of $N$ sites and magnetization $0<M_z<N/2$, the ground state, which
  contains $2M_z$ spinons, is reconfigured as the physical vacuum for a
  different species of quasi-particles, identifiable in the framework of the
  coordinate Bethe ansatz by characteristic configurations of Bethe quantum
  numbers. The dynamically dominant excitations are found to be scattering
  states of two such quasi-particles. For $N\to\infty$, these collective excitations
  form a continuum in $(q,\omega)$-space with an incommensurate soft mode. Their
  matrix elements in the dynamic spin structure factor $S_{zz}(q,\omega)$ are
  calculated directly from the Bethe wave functions for finite $N$. The
  resulting lineshape predictions for $N\to\infty$ complement the exact results
  previously derived via algebraic analysis for the exact 2-spinon part of
  $S_{zz}(q,\omega)$ in the zero-field limit. They are directly relevant for the
  interpretation of neutron scattering data measured in nonzero field on
  quasi-1D antiferromagnetic compounds.
\end{abstract}
\pacs{??}
\maketitle
%%%%%%%%%%%%%%%%%%%%%%%%%%%%%%%%%%%%%%%%%%%%%%%
%
\section{Introduction}\label{sec:I}
%
%%%%%%%%%%%%%%%%%%%%%%%%%%%%%%%%%%%%%%%%%%%%%%%
Advances in experimental techniques combined with improvements in sample preparation
make it possible to produce data of ever increasing resolution for the
quantum fluctuations and the underlying collective excitations in
quasi-one-dimensional (1D) magnetic compounds. Advances in the theoretical
analysis of relevant model systems combined with progress in the computational
treatment of aspects that remain elusive to exact analysis make it possible to
gain an ever more profound understanding of the observable collective
excitations in terms of a small number of constituent quasi-particles.

There is scarcely a better case for illustrating this multi-track advancement of
understanding quantum fluctuations than the 1D $s=\frac{1}{2}$ Heisenberg
antiferromagnet and the growing number of materials that have been discovered to
be physical realizations of this model system.  The Hamiltonian for $N$ spins
$\frac{1}{2}$ arranged in a cyclic chain with isotropic exchange coupling $J$
between nearest neighbors and a uniform magnetic field $h$,
\begin{equation}\label{eq:Hh}
H = \sum_{n=1}^N \left[J{\bf S}_n \cdot {\bf S}_{n+1}  - hS_n^z\right],
\end{equation}
is amenable to exact analysis via Bethe ansatz\cite{Beth31,FT81} and displays
dynamical properties of intriguing complexity. The field $h$ is a controllable
continuous parameter, which leaves the eigenvectors unaltered, but changes the
nature of the ground state via level crossings and thus has a strong impact on
the dynamical properties, in particular at low temperatures.

At $h\geq h_S=2J$ the ground state of $H$ has all spins aligned in field
direction: $|F\rangle\equiv|\uparrow\uparrow\cdots\uparrow\rangle$ is the reference state of the coordinate Bethe ansatz,
and all eigenstates are described as excitations of interacting magnons, a species
of spin-1 quasi-particles. Hence $|F\rangle$ plays the role of the magnon vacuum.
The ground state $|A\rangle$ of $H$ at $h=0$ contains $N/2$ magnons. The Bethe ansatz
enables us to reconfigure this state as the physical vacuum for a different
species of quasi-particles -- the spinons, which have spin $\frac{1}{2}$. The
entire spectrum of the Heisenberg model \eqref{eq:Hh} can also be generated as
composites of interacting spinons.\cite{note1}

Both descriptions are valid throughout the spectrum, but the magnon
interpretation is more useful near the magnon vacuum, and the spinon picture is
more useful near the spinon vacuum. The interaction energy of magnon scattering
states or spinon scattering states is of O($N^{-1}$) as long as the number of
quasi-particles in the collective excitations is of O(1).\cite{note2} In a
macroscopic system, the spectrum of such states is thus indistinguishable from
the corresponding free quasi-particle states. Even under these simplifying
circumstances, however, the interaction of the quasi-particles remains important
in the make-up of collective wave functions, and is likely to strongly affect
the transition rates and lineshapes.

At intermediate values $0<h<h_S$ of the magnetic field, the number of magnons
or spinons contained in the ground state of $H$ is of O($N$), implying that the
interaction energy for either quasi-particle species remains nonzero for
$N\to\infty$ in the ground state and in all low-lying excitations. This obscures
the role of individual magnons or spinons in the collective excitations and
obstructs the interpretation of spectral data obtained by experimental or
computational probes.

We can circumvent this problem by configuring the ground state $|G\rangle$ at
$0<M_z/N<\frac{1}{2}$ as the physical vacuum for yet a different species of
quasi-particles. From the new vantage point, the dynamically relevant collective
excitations are then again scattering states of few quasi-particles with an
interaction energy of O($N^{-1}$), which greatly facilitates the interpretation
of the spectra probed experimentally or computationally.

%%%%%%%%%%%%%%%%%%%%%%%%%%%%%%%%%%%%%%%%%%%%%%%
%
\section{Dynamic Structure Factor}\label{sec:II}
%
%%%%%%%%%%%%%%%%%%%%%%%%%%%%%%%%%%%%%%%%%%%%%%%
In an inelastic neutron scattering experiment performed at low temperature,
the observable scattering events predominantly involve transitions from the
ground state to a subset of collective excitations filtered from the rest
by selection rules and transition rates.  Under idealized circumstances, the
scattering cross section is proportional to the $T=0$ dynamic spin structure factor
\begin{equation}\label{eq:dssf}
S_{\mu\mu}(q,\omega) = 2\pi\sum_\lambda|\langle G|S_q^\mu|\lambda\rangle|^2\delta\left(\omega-\omega_\lambda \right),
\end{equation}
where $S_q^\mu = N^{-1/2}\sum_{n}e^{iqn}S_n^\mu$, $\mu=x,y,z$ is the spin fluctuation
operator.  In a macroscopic system, the aggregate of spectral lines in
\eqref{eq:dssf} pertaining to scattering events with energy transfer $\omega_\lambda\equiv
E_\lambda-E_G$, momentum transfer $q\equiv k_\lambda-k_G$, and transition rate $|\langle G|S_q^\mu|\lambda\rangle|^2$
form characteristic patterns of spectral weight in $(q,\omega)$-space. The shape of
the spectral weight distribution provides key information on how the dynamically
relevant collective excitations are composed of quasi-particles with specific
energy-momentum relations.

Experimentally it is possible, at least in principle, to separate the
information contained in the dynamic structure factors of the spin components
parallel and perpendicular to the field direction, i.e. the functions
$S_{zz}(q,\omega)$ and $S_{xx}(q,\omega) = \frac{1}{4}[S_{+-}(q,\omega) + S_{-+}(q,\omega)]$,
respectively, for the fluctuation operators $S_q^z$ and $S_q^{\pm}=S_q^x \pm
iS_q^y$. At $h=0$ the additional symmetry of $H$ dictates that $S_{zz}(q,\omega) =
\frac{1}{2}S_{+-}(q,\omega) = \frac{1}{2}S_{-+}(q,\omega)$.

\begin{figure}[t!]
\vspace*{-0.4cm}

\centerline{\epsfig{file=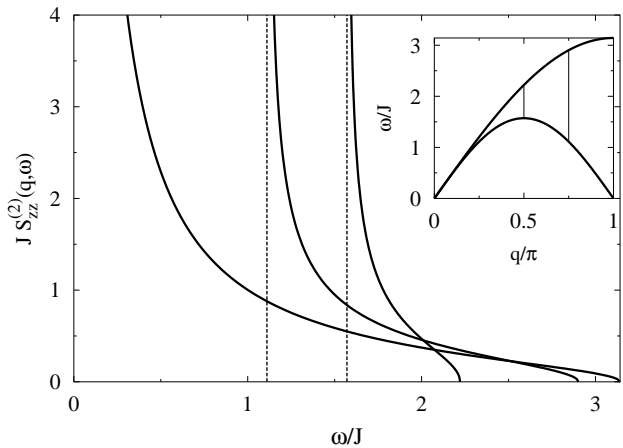,width=6.1cm,angle=-90}}
\caption{Exact 2-spinon lineshapes at $q=\pi/2,3\pi/4,\pi$ of $S_{zz}(q,\omega)$ at $T=0$ for
  Hamiltonian \eqref{eq:Hh} at $h=0$ as determined via algebraic
  analysis. The inset shows the boundaries \eqref{eq:epslu} of the 2-spinon
  spectrum.}
\label{fig:1}
\end{figure}

An anchor point for the new results presented in the following is the
exact 2-spinon dynamic spin structure factor at $T=0$, which was determined
recently via algebraic analysis and shown to contribute 73\% of the total
intensity in $S_{zz}(q,\omega)$ at $h=0$.\cite{KMB+97} Given the
energy-momentum relation\cite{FT81}
\begin{equation}\label{eq:spdisp}
\epsilon_{sp}(p) = \frac{\pi}{2}J\sin p,\quad 0\leq p\leq\pi,
\end{equation}
of the spinon quasi-particle, the 2-spinon states with wave numbers $q=p_1+p_2$
and energy $\omega=\epsilon_{sp}(p_1)+\epsilon_{sp}(p_2)$ form a continuum confined by the
boundaries\cite{DP62,Yama69}
\begin{equation}\label{eq:epslu}
\epsilon_L(q) = \frac{\pi}{2}J|\sin q|, \quad \epsilon_U(q) = \pi
J\left|\sin\frac{q}{2}\right|,
\end{equation}
as illustrated in Fig.~\ref{fig:1} (inset). The main plot shows the exact
2-spinon lineshapes of $S_{zz}(q,\omega)$ at $q=\pi/2,3\pi/4,\pi$.  The most detailed
experimental data available for testing these results pertain to
KCuF$_3$.\cite{TCNT95}

We shall see that the magnetic field causes dramatic
changes in both the spectrum and the lineshapes. At the root of these changes is
a change in the nature of the relevant quasi-particles. Two compounds suitable for
studying magnetic-field effects on spectrum and lineshapes are
Cu(C$_6$D$_5$COO)$_2\cdot$3D$_2$O and
Cu(C$_4$H$_4$N$_2$)(NO$_3$)$_2$.\cite{DHR+97,HSR+99}

%%%%%%%%%%%%%%%%%%%%%%%%%%%%%%%BEGIN-FIGURE%%%%
%%%%%%%%%%%%%%%%%%%%%%%%%%%%%%%%END-FIGURE%%%%%

%%%%%%%%%%%%%%%%%%%%%%%%%%%%%%%%%%%%%%%%%%%%%%%
%
\section{Bethe Ansatz Equations}\label{sec:III}
%  
%%%%%%%%%%%%%%%%%%%%%%%%%%%%%%%%%%%%%%%%%%%%%%%
The Bethe ansatz\cite{Beth31} is an exact method for the calculation of
eigenvectors of integrable quantum many-body systems. The Bethe wave function of
any eigenstate of \eqref{eq:Hh} in the invariant subspace with $r=N/2-M_z$
reversed spins relative to the magnon vacuum,
\begin{equation}\label{eq:psir}
 |\psi\rangle = \sum_{1\leq n_1<\ldots<n_r\leq N} a(n_1,\ldots,n_r)
 S_{n_1}^-\cdots S_{n_r}^-|F\rangle,
\end{equation}
has coefficients of the form
\begin{eqnarray}\label{eq:bar}\hspace*{-0.4cm}
 a(n_1,\ldots,n_r) =
 \!\sum_{{\cal P}\in S_r}
 \!\exp && \left(\!i\sum_{j=1}^r k_{{\cal P} j}n_j
 + \frac{i}{2}\sum_{i<j}^{r} \theta_{{\cal P}i{\cal P}j}\!\right)
\end{eqnarray}
determined by $r$ magnon momenta $k_i$ and one phase angle
$\theta_{ij}=-\theta_{ji}$ for each magnon pair. The sum ${\cal P}\in S_r$ is
over the permutations of the labels $\{1,2,\ldots,r\}$. 

The consistency requirements for the coefficients $a(n_1,\ldots,n_r)$ inferred from
the eigenvalue equation $H|\psi\rangle=E|\psi\rangle$ and the requirements imposed
by translational invariance lead to a set of coupled nonlinear
equations for the $k_i$ and $\theta_{ij}$. A computationally
convenient rendition of the Bethe ansatz equations has the form 
\begin{equation}
\label{eq:bae}
N\phi(z_i) = 2\pi I_i + \sum_{j\neq i}\phi\bigl [(z_i-z_j)/2\bigr ],
\quad i=1,\ldots,r,
\end{equation}
where $\phi(z) \equiv 2\arctan z$, $k_i = \pi -\phi(z_i)$ and $\theta_{ij} = \pi \, {\rm
  sgn}[{\Re}(z_i-z_j)] - \phi\bigl[(z_i-z_j)/2\bigr]$.  Every solution of
\eqref{eq:bae} is specified by a set of Bethe quantum numbers
$I_1<I_2<\cdots<I_r$, which assume integer values for odd $r$ and half-integer values
for even $r$. The energy and wave number of the eigenvector thus determined
are
\begin{equation}\label{eq:ekz}
\frac{E-E_F}{J} = -\sum_{i=1}^{r}\frac{2}{1+z_i^2},\quad
k = \pi r - \frac{2\pi}{N}\sum_{i=1}^r I_i,
\end{equation}
where $E_F=JN/4$ is the energy of the magnon vacuum.

We consider the class $K_r$ of eigenstates whose Bethe quantum numbers comprise,
for $0\leq r\leq N/2$ and $0\leq m\leq N/2-r$, all configurations
\begin{equation}\label{eq:I2msp}
-\frac{r}{2} + \frac{1}{2} -m \leq I_1 < I_2 < \cdots < I_r \leq \frac{r}{2}
- \frac{1}{2} + m.
\end{equation}

Here we employ the solutions $\{z_i\}$ of the Bethe ansatz equations not only to
generate spectral data via \eqref{eq:ekz}, which is standard practice, but also
to evaluate transition rates $|\langle G|S_q^\mu|\lambda\rangle|^2$ for the dynamic structure factor
\eqref{eq:spdisp} directly from the normalized Bethe wave functions
$|\lambda\rangle\equiv|\psi\rangle/||\psi||$. The computational aspects of this method are discussed
elsewhere.\cite{KHM00}

%%%%%%%%%%%%%%%%%%%%%%%%%%%%%%%%%%%%%%%%%%%%%%%
%
\section{Physical Vacuum and Quasi-Particles}\label{sec:IV}
%  
%%%%%%%%%%%%%%%%%%%%%%%%%%%%%%%%%%%%%%%%%%%%%%%
The ground-state wave function $|G\rangle$ at $0\leq M_z\leq N/2$ is specified by the set of
$r=N/2-M_z$ Bethe quantum numbers\cite{YY66a}
\begin{equation}\label{eq:IG}
\{I_i\}_G = \left\{-\frac{N}{4}+\frac{M_z}{2}+\frac{1}{2},\ldots,
\frac{N}{4}-\frac{M_z}{2}-\frac{1}{2} \right\}.
\end{equation}  
As the magnetic field increases from $h=0$ to $h_S=2J$, the magnetization $M_z$
increases in units of one from zero to $N/2$. A sequence of level crossings
produces a magnetization curve $(M_z/N$ versus $h)$ in the form of a staircase
with $N/2$ steps of height $1/N$, which converges toward a smooth line as
$N\to\infty$.\cite{Grif64,BF64,KHM98}

\begin{figure}[t!]
\vspace*{0.3cm}

\centerline{\epsfig{file=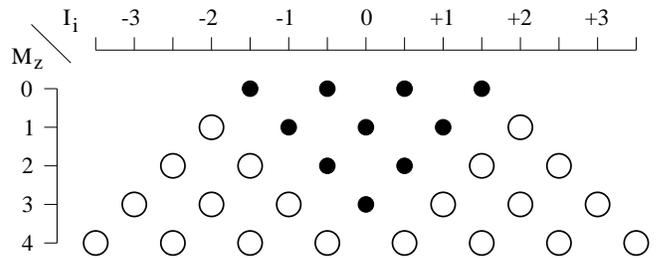,width=8.5cm}}
\caption{Physical vacuum $|G\rangle$ for a chain of $N=8$ spins at magnetization
  $M_z=0,1,\ldots,4$. The values of the Bethe quantum numbers $I_i$ are given by the
  positions of the magnons ({\Large $\bullet$}). The spinons ($\bigcirc$) correspond to
  vacancies in the $I_i$-configurations.  }

\label{fig:2}
\end{figure}

Depending on the reference state used for the characterization of the ground
state $|G\rangle$, it can be regarded as a scattering state of $N/2-M_z$ magnons
excited from the magnon vacuum $|F\rangle$ or as a scattering state of $2M_z$ spinons
excited from the spinon vacuum $|A\rangle$. To illustrate the distinct roles played by
the two species of quasi-particles in the class-$K_r$ states, we show in
Fig.~\ref{fig:2} the configuration of Bethe quantum numbers for $|G\rangle$ in a
system with $N=8$ and all values of $M_z$ realized between $h=0$ and $h=h_S$.
The positions of the magnons ({\Large $\bullet$}) are determined by the set
\eqref{eq:IG} of $I_i$'s and the positions of the spinons ($\bigcirc$) by the vacancies
across the full range of the $I_i$'s allowed by \eqref{eq:I2msp} for class $K_r$
states.

%%%%%%%%%%%%%%%%%%%%%%%%%%%%%%%BEGIN-FIGURE%%%%
%%%%%%%%%%%%%%%%%%%%%%%%%%%%%%%%END-FIGURE%%%%%

Henceforth we treat $|G\rangle$ as the new physical vacuum. At $h=0$ (top row in
Fig.~\ref{fig:2}) it coincides with the spinon vacuum, a state with $N/2$
magnons. At $h=h_S$ (bottom row) it coincides with the magnon vacuum, a state
containing $N$ spinons.  All states within class $K_r$ are generated from $|G\rangle$
by rearranging the magnons or (equivalently) the spinons into all allowed
configurations.

For $r=N/2$ (top row) and $r=0$ (bottom row) the state shown is the only
possible configuration within class $K_r$.  In the fourth row, the lone magnon
can be moved across the array of spinons, generating a branch (one-parameter
set) of 1-magnon excitations for $N\to\infty$. In the second row, the two spinons can
be moved independently across the array of magnons, generating a continuum
(two-parameter set) of 2-spinon excitations for $N\to\infty$ with boundaries
\eqref{eq:epslu} as shown in Fig.~\ref{fig:1}.  The center row in
Fig.~\ref{fig:2} pertains to the field at half the saturation magnetization
$(M_z=r=N/4)$, the case we shall investigate extensively for various system
sizes. Here $|G\rangle$ contains twice as many spinons as it contains magnons.

The integer $m$ with range $0<m\leq M_z$ used in \eqref{eq:I2msp} is a convenient
quantum number for the subdivision of the classes $K_r$. Every state of $K_r$ at
fixed $m$ can then be regarded as a scattering state of $m$ pairs of spinon-like
quasi-particles. To distinguish them from the spinons, we name the new
quasi-particles {\em psinons}.

The ground state $|G\rangle$, the only state with $m=0$, is the psinon vacuum. Here
the magnons form a single array flanked by two arrays of spinons (see
Fig.~\ref{fig:2}).  Relaxing the constraint in (\ref{eq:I2msp}) from $m=0$ to
$m=1$ yields a two-parameter set of states -- the 2-psinon excitations. Here the
array of magnons breaks into three clusters separated by the two innermost
spinons, which now assume the role of psinons. The remaining $2M_z-2$ spinons
stay sidelined.  In the 4-psinon states $(m=2)$, two additional spinons have
been mobilized into psinons. By this prescription, we can systematically
generate sets of $2m$-psinon excitations for $0\leq m\leq M_z$.

\begin{figure}[tb]\vspace*{-0.3cm}
\includegraphics[width=7.2cm,angle=-90]{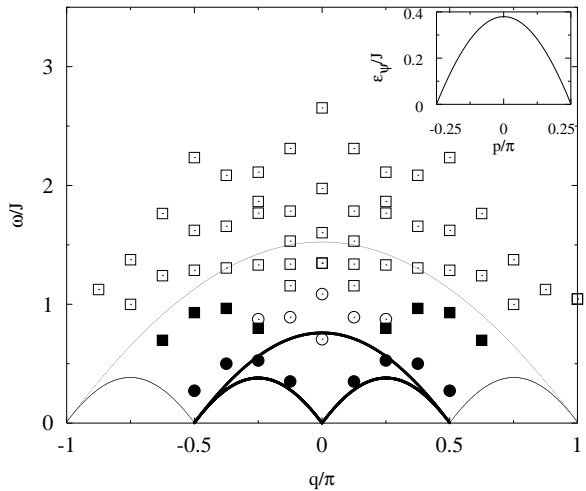}
\caption{Spectrum of 2-psinon excitations (circles) and
  4-psinon excitations (squares) for $M_z=N/4$ and $N=16$. The states marked by
  full symbols are dynamically dominant in $S_{zz}(q,\omega)$. The spectral ranges of
  the 2-psinon states (thick lines) and 4-psinon states (thin lines) for $N\to\infty$
  are inferred from data for $N=2048$. The inset shows the psinon energy
  momentum relation $\epsilon_\psi(p)$.}
\label{fig:3}
\end{figure}

To illustrate the quasi-particle role of the psinons in the class-$K_r$
collective states we have plotted in Fig.~\ref{fig:3} energy versus wave number
of all 2-psinon states (circles) and 4-psinon states (squares) at $M_z=N/4$ for
$N=16$. Also shown are the spectral boundaries of 2-psinon and
4-psinon states for $N\to\infty$ as inferred from solutions of \eqref{eq:bae} for
$N=2048$. The 2-psinon continuum, outlined by thick lines, is confined to the
interval $|q|\leq q_s$, where
\begin{equation}\label{eq:qs}
 q_s \equiv \pi(1-2M_z/N)
\end{equation}
denotes the wave number of an incommensurate soft mode. The lower 4-psinon
spectral boundary is the same as the 2-psinon lower boundary but extended
periodically over the entire Brillouin zone. The upper 4-psinon boundary is
related to the upper 2-psinon boundary by a scale transformation $(q\to 2q,\omega\to
2\omega)$.

%%%%%%%%%%%%%%%%%%%%%%%%%%%%%%%BEGIN-FIGURE%%%%
%%%%%%%%%%%%%%%%%%%%%%%%%%%%%%%%END-FIGURE%%%%%

The relationship between the ranges in $(q,\omega)$-space of the 2-psinon states and
the 4-psinon states does indeed reflect the fact that they are scattering states
of two or four quasi-particles, respectively, of the same species. Like the
spinon, the psinon is not observable in isolation via neutron scattering, but
its energy momentum relation $\epsilon_\psi(p), -\pi/4\leq p\leq\pi/4$, can be inferred from the
data of Fig.~\ref{fig:3} (see inset).

If there were no psinon interaction, the wave number and energy of a $2m$-psinon
state would be $q = \sum_{i=1}^{2m}p_i$, $\omega = \sum_{i=1}^{2m}\epsilon_\psi(p_i)$. The $N=16$
data make it quite clear that the finite-size energy correction caused by the
psinon interaction is stronger in the 4-psinon states than in the 2-psinon
states. In both sets of collective states, the interaction energy goes to zero
as the scattering events become less and less frequent in a chain of increasing
length.  However, it takes longer chains for finite-$N$ 4-psinon data to reach
comparable convergence toward the spectral boundaries predicted for $N\to\infty$,
because for fixed $N$, the scattering events between psinons are more numerous
in a typical 4-psinon state than in a typical 2-psinon state.

If instead of the psinon vacuum we had used the spinon vacuum as the reference
state at $M_z=N/4$, then both the 2-psinon states and the 4-psinon states would
have to be described as scattering states of $N/2$ spinons.  Although we know
the energy-momentum relation of a spinon, Eq.  \eqref{eq:spdisp}, it is of
little use to determine the spectral threshold in Fig.~\ref{fig:3}. Since the
2-psinon and 4-psinon states maintain a finite density of spinons in the limit
$N\to\infty$, the spinon interaction energy remains significant.  This problem does not
arise at $M_z=0$. In the 2-spinon scattering states depicted in
Fig.~\ref{fig:1}, the spinon interaction energy vanishes for $N\to\infty$ just as the
psinon interaction energy does in the 2-psinon and 4-psinon scattering states
depicted in Fig.~\ref{fig:3}.\cite{note7}

%%%%%%%%%%%%%%%%%%%%%%%%%%%%%%%%%%%%%%%%%%%%%%%
%
\section{ Dynamically relevant excitations}\label{sec:V}
%  
%%%%%%%%%%%%%%%%%%%%%%%%%%%%%%%%%%%%%%%%%%%%%%%
At $M_z=0$ the spectral weight in the dynamic spin structure factor
$S_{zz}(q,\omega)$ is dominated by the 2-spinon excitations.\cite{KMB+97} Our task
here is to determine how the spectral weight of $S_{zz}(q,\omega)$ at $M_z\neq 0$ is
distributed among the $2m$-psinon excitations. In investigating this question,
we follow the strategy of an older study\cite{MTBB81} but with vastly improved
conceptual and numerical tools.

\begin{figure}[bt]
  \centerline{\epsfig{file=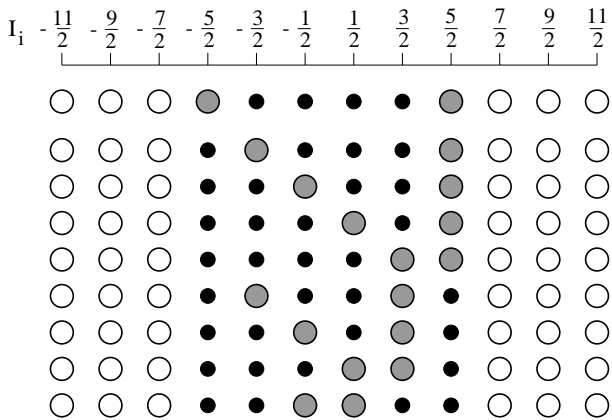,width=8.0cm}}

\caption{Psinon vacuum $|G\rangle$ for $N=16, M_z=4$ and 2-psinon states
  with $q\geq0$. The $I_i$ values are marked by the positions of the magnons (small
  circles). The spinons (large circles) mark $I_i$-vacancies. A subset of the
  spinons are called psinons (grey circles).  }
\label{fig:4}
\end{figure}

We begin by exploring, in a chain of $N=16$ spins at $M_z/N=\frac{1}{4}$, the
transition rates between the ground state $|G\rangle$ and all $2m$-psinon excitations
for $m=0,1,2,3,4$. The Bethe quantum numbers of the states with $m=0,1$ are
shown in Fig.~\ref{fig:4}. The first row represents the psinon vacuum with its
four magnons sandwiched by two sets of four spinons. The two innermost spinons
(marked grey) become psinons when at least one of them is moved to another
position.  In the rows underneath, the psinons are moved systematically across
the array of magnons while the remaining spinons stay frozen in place. These
eight configurations describe all 2-psinon states with $q\geq 0$.

%%%%%%%%%%%%%%%%%%%%%%%%%%%%%%%BEGIN-FIGURE%%%%
%%%%%%%%%%%%%%%%%%%%%%%%%%%%%%%%END-FIGURE%%%%%

The wave numbers, energies, and transition rates of the states shown in
Fig.~\ref{fig:4} are listed in Table~\ref{tab:I}.  Remarkably, almost the entire
2-psinon spectral weight is concentrated in the lowest excitation for any given
$q$. The dynamically dominant 2-psinon states are marked by solid circles in
Fig.~\ref{fig:3}. In a macroscopic system, they form the lower boundary of the
2-psinon continuum.

%%%%%%%%%%%%%%%%%%%%%%%%%%%%BEGIN-TABLE%%%%%%%%
\begin{table}[bt]
\caption{Ground state $|G\rangle$ and 2-psinon excitations for $N=16$, $M_z=4$, and wave
  numbers $q\equiv k-k_G\geq 0$ (in units of $2\pi/N$). The ground state has
  $k_G=0$ and $E_G=-11.5121346862$.}
\begin{center}
  \begin{tabular}[h]{cccc}\hline\hline
$2I_i$ & $k-k_G$ & $E-E_G$ & $|\langle G|S_q^z|\lambda\rangle|^2$ \\ \hline
$-3-1+1+3$ & 0 & 0.0000000000 & 1.0000000000 \\
$-5-1+1+3$ & 1 & 0.3504534152 & 0.0484825989 \\
$-5-3+1+3$ & 2 & 0.5271937189 & 0.0587154211 \\
$-5-3-1+3$ & 3 & 0.5002699273 & 0.0773592284 \\
$-5-3-1+1$ & 4 & 0.2722787522 & 0.1257902349 \\
$-5-1+1+5$ & 0 & 0.7060324808 & 0.0000000000 \\
$-5-3+1+5$ & 1 & 0.8908215652 & 0.0000064288 \\
$-5-3-1+5$ & 2 & 0.8738923064 & 0.0000312622 \\
$-5-3+3+5$ & 0 & 1.0855897189 & 0.0000000000 \\
\hline\hline
  \end{tabular}
\end{center}
\label{tab:I}
\end{table}
%%%%%%%%%%%%%%%%%%%%%%%%%%%%%%%%END-TABLE%%%%%%

Next we calculate the transition rates $|\langle G|S_q^z|\lambda\rangle|^2$ for the complete set
of 4-psinon states.  Interestingly, we observe that most of the 4-psinon
spectral weight is again carried by a single branch of excitations. The
dynamically dominant 4-psinon states for $N=16$ are shown as full squares in
Fig.~\ref{fig:3}. For large $N$ they form a branch adjacent to the 2-psinon
spectral threshold.

\begin{figure}[tb]
\centerline{\epsfig{file=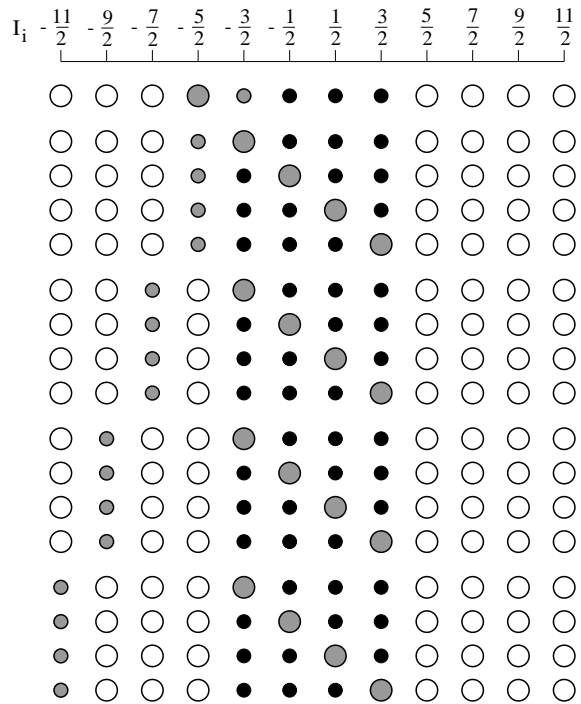,width=7.5cm}}

\caption{Psinon vacuum $|G\rangle$ for $N=16,M_z=4$ and set of $\psi\psi^*$ states with $0\leq q\leq \pi$.
  The $I_i$ are given by the positions of the magnons (small circles) in each
  row. The spinons (large circles) correspond to $I_i$-vacancies. The psinon
  $(\psi)$ and the antipsinon $(\psi^*)$ are marked by a large and a small grey
  circle, respectively. }
\label{fig:5}
\end{figure}

An investigation of the remaining $2m$-psinon states shows that there exists one
dynamically dominant branch of $2m$-psinon excitations for $0<m\leq M_z$. The
configurations of Bethe quantum numbers pertaining to the four branches for
$N=16$, each consisting of $N/2-M_z=4$ states (at $q>0$), are shown in
Fig.~\ref{fig:5}. The energies, wave numbers, and transition rates of these
excitations are listed in Table~\ref{tab:II}. All other $2m$-psinon excitations
have transition rates that are smaller by at least two orders of magnitude at
$q<\pi/2$, and still by more than one order of magnitude at $q\geq\pi/2$.

%%%%%%%%%%%%%%%%%%%%%%%%%%%%%%%BEGIN-FIGURE%%%%
%%%%%%%%%%%%%%%%%%%%%%%%%%%%%%%%END-FIGURE%%%%%

%%%%%%%%%%%%%%%%%%%%%%%%%%%%BEGIN-TABLE%%%%%%%%
\begin{table}[tb]
\caption{Ground state and dynamically dominant excitations for ($N=16$, $r=4$) among 
  $2m$-psinon states ($m=0,1,\ldots,4$). The latter form the $\psi\psi^*$ continuum in the 
  limit $N\to\infty$. The wave numbers $q\equiv k-k_G\geq 0$ are in units of
  $2\pi/N$.}
\begin{center}
  \begin{tabular}[h]{ccccc}\hline\hline
$2I_i$ & $2m$ & $q$ & $E-E_G$ & $|\langle G|S_q^z|\lambda\rangle|^2$ \\ \hline
$-3-1+1+3$ & 0 & 0 & 0.0000000000 & 1.0000000000 \\
$-5-1+1+3$ & 2 & 1 & 0.3504534152 & 0.0484825989 \\
$-5-3+1+3$ & 2 & 2 & 0.5271937189 & 0.0587154211 \\
$-5-3-1+3$ & 2 & 3 & 0.5002699273 & 0.0773592284 \\
$-5-3-1+1$ & 2 & 4 & 0.2722787522 & 0.1257902349 \\
$-7-1+1+3$ & 4 & 2 & 0.7981588810 & 0.0426892576 \\
$-7-3+1+3$ & 4 & 3 & 0.9653287066 & 0.0552255878 \\
$-7-3-1+3$ & 4 & 4 & 0.9301340415 & 0.0743667351 \\
$-7-3-1+1$ & 4 & 5 & 0.6966798553 & 0.1253357676 \\
$-9-1+1+3$ & 6 & 3 & 1.2708459328 & 0.0345439774 \\
$-9-3+1+3$ & 6 & 4 & 1.4285177129 & 0.0516860817 \\
$-9-3-1+3$ & 6 & 5 & 1.3858078992 & 0.0753564030 \\
$-9-3-1+1$ & 6 & 6 & 1.1488426600 & 0.1406415212 \\
$-11-1+1+3$ & 8 & 4 & 1.6819046570 & 0.0235815843 \\
$-11-3+1+3$ & 8 & 5 & 1.8257803105 & 0.0443726010 \\
$-11-3-1+3$ & 8 & 6 & 1.7724601200 & 0.0744641955 \\
$-11-3-1+1$ & 8 & 7 & 1.5309413164 & 0.1686893882 \\
\hline\hline
  \end{tabular}
\end{center}
\label{tab:II}
\end{table}
%%%%%%%%%%%%%%%%%%%%%%%%%%%%%%%%END-TABLE%%%%%%

Inspection of Fig.~\ref{fig:5} reveals an interesting pattern, indicative of the
composition of the dynamically relevant collective excitations. They form a
two-parameter set. The two parameters are highlighted by grey circles.  Hitherto
we have interpreted each group of four configurations as a branch of $2m$-psinon
excitations, which are seemingly arbitrary one-parameter subsets taken from
$2m$-parameter sets of states. In a macroscopic system, all but the lowest such
branches contain a macroscopic number of psinons. Hence the range of the
dynamically relevant excitations in $(q,\omega)$-space cannot be inferred from the
psinon energy-momentum relation alone as was possible for the 2-psinon and
4-psinon continua, because the psinon interaction energy will remain
non-negligible in most of these states for $N\to\infty$, just as the spinon interaction
energy was non-negligible in the 2-psinon and 4-psinon scattering states at
$M_z\neq 0$.

A more natural interpretation of the pattern on display in Fig.~\ref{fig:5}
identifies one of the two parameters as a psinon (large grey circle) as before
and the other parameter as a new quasi-particle (small grey circle). The latter
is represented by a hole in what was one of two spinon arrays in the psinon
vacuum. Instead of focusing on the cascade of psinons (mobile spinons) which
this hole has knocked out of the vacuum, we focus on the hole itself, which has
properties commonly attributed to antiparticles. The psinon $(\psi)$ and the
antipsinon $(\psi^*)$ exist in disjunct parts of the psinon vacuum, namely in the
magnon and spinon arrays, respectively. When they meet at the border of the two
arrays, they undergo a mutual annihilation, represented by the step from the
second row to the top row in Fig.~\ref{fig:5}.

We could have interpreted the small grey circle as a magnon (spin-1
quasi-particle), but when we do that we must take into account that it then
coexists in the magnon vacuum with a macroscopic number of fellow magnons (small
black circles). From this perspective, the collective excitation must be viewed
as containing a finite density of magnons (for $N\to\infty$), in which the magnon
interaction remains energetically significant for scattering states.  The
nonzero interaction energy obscures the role of individual magnons.

On the other hand, when the small grey circle is interpreted as an antipsinon,
then it lives in the psinon vacuum, i.e. almost in isolation. The only other
particle present is a psinon (large grey circle). In a macroscopic system, the
interaction energy in a psinon-antipsinon $(\psi\psi^*)$ scattering state becomes
negligible. Therefore, the identity of both quasi-particles is easily
recognizable in the spectrum.

\begin{figure}[tb]
  
  \centerline{\epsfig{file=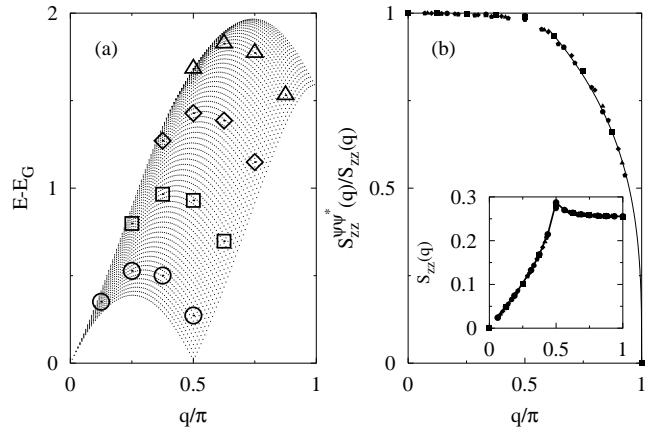,width=6.7cm,angle=-90}}
\caption{(a)  $\psi\psi^*$ excitations at $M_{z}/N=1/4$ for 
  $N=16$ (circles, squares, diamonds, triangles for $m=1,2,3,4$,
  respectively) and $N=256$ (dots). (b) Integrated intensity $S_{zz}(q)$ (inset)
  and relative $\psi\psi^*$ contribution (main plot) for $N=12, 16, 20, 24, 28, 32$.
  The lines connect the $N=32$ data points.}
\label{fig:6}
\end{figure}

The energies versus the wave numbers of the 16 $\psi\psi^*$ states listed in Table
\ref{tab:II} are shown in Fig.~\ref{fig:6}(a) as large symbols. The four
branches from bottom to top pertain to $m=1,\ldots,4$. Also shown in the same plot
are the $\psi\psi^*$ states for $N=256$. The lower boundary of the $\psi\psi^*$continuum
emerging in the limit $N\to\infty$ touches down to zero frequency at $q=0$ and $q =
q_s=\pi/2$. Between $q_s$ and $\pi$, it rises monotonically and reaches the value
$E-E_G = h$. A direct observation of the incommensurate soft mode at $q_s$ was
made in a neutron scattering experiment on Cu(C$_6$D$_5$COO)$_2\cdot$3D$_2$O (copper
benzoate).\cite{DHR+97}

%%%%%%%%%%%%%%%%%%%%%%%%%%%%%%%BEGIN-FIGURE%%%%
%%%%%%%%%%%%%%%%%%%%%%%%%%%%%%%%END-FIGURE%%%%%

Figure~\ref{fig:6}(b) shows the relative integrated intensity of the $\psi\psi^*$
excitations for various $N$ at fixed $M_z/N=\frac{1}{4}$. At $q\lesssim q_s=\pi/2$, virtually
all spectral weight of $S_{zz}(q,\omega)$ originates from $\psi\psi^*$ fluctuations. An
extrapolation of the data points at $q=\pi/2$ suggests that the relative
$\psi\psi^*$ spectral weight is in excess of 93\%.

At $q\gtrsim q_s$ the $\psi\psi^*$ contribution to the integrated intensity decreases
monotonically but stays dominant over more than half the distance to the zone
boundary. The width of the $\psi\psi^*$ continuum vanishes linearly on approach of
$q=\pi$, and the relative spectral weight more slowly: $S_{zz}(q)\sim (\pi-q)^\gamma$, $\gamma\simeq
0.3$.  This enhances the observability of the $\psi\psi^*$ excitations in the narrow
energy range near the Brillouin zone in spite of the low absolute intensity.
Finite-$N$ data for the integrated intensity $S_{zz}(q)$ are shown in the inset
to Fig.~\ref{fig:6}(b).  This function is peaked at $q=q_s$, where the $\psi\psi^*$
spectral weight is overwhelmingly predominant.

When we lower $M_z$, the soft mode at $q_s$ moves to the right, the number of
$2m$-psinon branches that contribute to the $\psi\psi^*$ continuum shrinks but each
branch gains additional states. At $M_z=1$ we are left with one 2-psinon branch
extending over the interior of the entire Brillouin zone. This branch is equal
to the lowest branch of 2-spinon states with dispersion $\epsilon_L(q)$, Eq.
\eqref{eq:epslu}. However, even for this case the psinon vacuum is different
from the spinon vacuum. The former is the lowest-energy 2-spinon state (with
$M_z=1$), whereas the latter is a state with $M_z=0$. The wave number of the two
vacua differ by $\pi$. At $M_z=0$ the $\psi\psi^*$ excitations disappear altogether.
The limit $h\to 0$ of the infinite chain is very subtle and will be discussed
elsewhere.\cite{note3}

When we increase $M_z$ toward the saturation value, the soft mode moves to the
left, and the number of $2m$-psinon branches increases, but each
branch becomes shorter. At $M_z=N/2-1$, the two-parameter set collapses
into a one-parameter set consisting of one $2m$-psinon state each for
$m=1,2,\ldots,N/2-1$. These states are more naturally interpreted as a branch of
1-magnon excitations with dispersion $\epsilon_1(q)=J(1-\cos q)$. Their relative
spectral weight in $S_{zz}(q,\omega)$ is now 100\%, but the absolute intensity for
$q\neq0$ is only of O($N^{-1}$).

\begin{figure}[tb]\vspace*{-0.3cm}
  \centerline{\epsfig{file=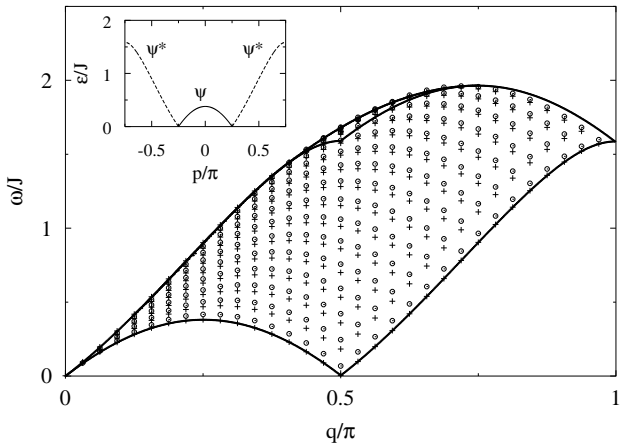,width=7.2cm,angle=-90}}

\caption{Energy versus wave number of all $\psi\psi^*$ scattering states at $q\geq0$ for
  $N=64$ ($\circ$) in comparison with the corresponding free $\psi\psi^*$ states
  $(+)$. The inset shows the energy-momentum relations of the psinon $(0\leq |p|\leq
  \pi/4)$ and the antipsinon $(\pi/4\leq |p|\leq3\pi/4)$ as inferred from from $\psi\psi^*$ data
  for $N=2048$.}
\label{fig:7}
\end{figure}

To further illustrate the roles of the psinon and the antipsinon as the relevant
quasi-particles in the collective excitations dominating the spectral weight in
$S_{zz}(q,\omega)$, we compare in Fig.~\ref{fig:7} the energies between the $\psi\psi^*$
scattering states for $N=64$ and the corresponding (fictitious) free $\psi\psi^*$
superpositions. The vertical displacement of any $(\circ)$ from the associated
$(+)$ reflects the interaction energy between the two quasi-particles. This
energy approaches zero for all states of this class as $N\to\infty$.

The energy-momentum relations of the two quasi-particles can be accurately
inferred from $N=2048$ data for the spectral thresholds of the $\psi\psi^*$ states as
illustrated in the inset to Fig.  \ref{fig:7}.  The psinon dispersion $\epsilon_\psi(p)$
is confined to the interval at $0\leq |p|\leq \pi/4$ (solid line) and the antipsinon
dispersion $\epsilon_{\psi^*}(p)$ to $\pi/4\leq |p|\leq3\pi/4$ (dashed line). The different ranges
of momentum which the two quasi-particles are allowed to have correspond to the
different regions in Fig.  \ref{fig:5} across which the circles pertaining to
them can be varied.

%%%%%%%%%%%%%%%%%%%%%%%%%%%%%%%BEGIN-FIGURE%%%%
%%%%%%%%%%%%%%%%%%%%%%%%%%%%%%%%END-FIGURE%%%%%

The lower boundary of the $\psi\psi^*$ continuum is defined by collective states in
which one of the two particles has zero energy: the psinon for $0\leq |q|\leq \pi/2$
and the antipsinon for $\pi/2\leq |q|\leq\pi$. The upper boundary consists of three
distinct segments. 

For $0\leq q\lesssim 0.3935$ the highest-energy $\psi\psi^*$ state is made up of a zero-energy
psinon with momentum $p_\psi=-\pi/4$ and an antipsinon with momentum $p_{\psi^*}=\pi/4+q$.
Here the shape of the continuum boundary is that of the psinon dispersion.
Likewise, for $3\pi/4\leq q\leq\pi$, the states along the upper continuum boundary are
made up of a maximum-energy antipsinon (with momentum $p_{\psi^*}=3\pi/4$ and a
psinon with momentum $p_\psi=-3\pi/4+q$. Here the shape of the continuum boundary is
that of the psinon dispersion.

When these two delimiting curves are extended into the middle segment, $0.3935\lesssim
q\leq3\pi/4$, they join in a cusp singularity at $q=\pi/2$. Here the highest $\psi\psi^*$
state does not involve any zero-energy quasi-particles. The maximum of
$\epsilon_\psi(p_\psi)+\epsilon_{\psi^*}(p_{\psi^*})$ subject to the constraint $p_\psi+p_{\psi^*}=q$ does not
occur at the endpoint of any quasi-particle dispersion curve. Consequently, the
$\psi\psi^*$ continuum is partially folded about the upper continuum boundary along
the middle segment.

%%%%%%%%%%%%%%%%%%%%%%%%%%%%%%%%%%%%%%%%%%%%%%%
%
\section{ Lineshapes}\label{sec:VI}
%  
%%%%%%%%%%%%%%%%%%%%%%%%%%%%%%%%%%%%%%%%%%%%%%%
To calculate the lineshapes relevant for fixed-$q$ scans in an inelastic neutron
scattering experiment from the spectrum and matrix elements obtained via Bethe
ansatz, we exploit key properties of transition rates and densities of states of
sets of excitations that form two-parameter continua in $(q,\omega)$ space for $N\to\infty$.
The $\psi\psi^*$ transition rates (scaled by $N$) form a continuous function
$M_{zz}^{\psi\psi^*}(q,\omega)$ for $N\to\infty$.  The $\psi\psi^*$ density of states (scaled by
$N^{-1}$) becomes a continuous function $D^{\psi\psi^*}(q,\omega)$ for $N\to\infty$.  The $\psi\psi^*$
spectral-weight distribution is then the product $S_{zz}^{\psi\psi^*}(q,\omega)
=D^{\psi\psi^*}(q,\omega)M_{zz}^{\psi\psi^*}(q,\omega)$.\cite{note4} In the following, we consider
three wave numbers at $M_z=N/4$.

\begin{figure}[tb]
  \centerline{\epsfig{file=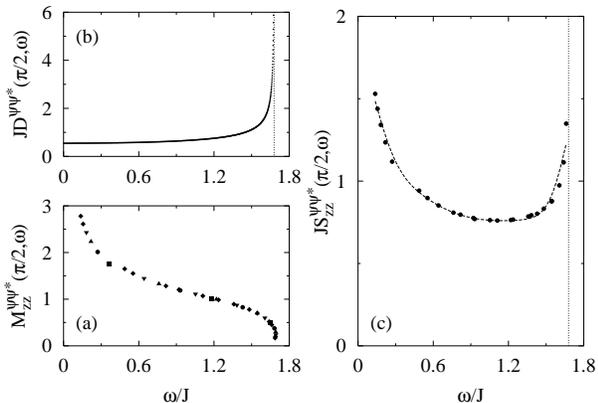,width=6.2cm,angle=-90}}

\caption{(a) Density of $\psi\psi^*$ states at $q=\pi/2$ evaluated via
  \eqref{eq:Dpsi} from Bethe ansatz data for $N=2048$. (b) Transition rates
  \eqref{eq:Mpsi} between the psinon vacuum and the $\psi\psi^*$ states at $q=\pi/2$ for
  $N=12, 16, 20, 24, 28, 32$. (c) Lineshape at $q=\pi/2$ of the $\psi\psi^*$
  contribution to $S_{zz}(q,\omega)$. All results are for $M_z=N/4$.}
\label{fig:8}
\end{figure}

At $q=\pi/2$, the $\psi\psi^*$ continuum is gapless and the relative $\psi\psi^*$ spectral
weight in $S_{zz}(q,\omega)$ has a maximum. The scaled density of $\psi\psi^*$ states is
generated from $N=2048$ data of the set of points
\begin{equation}\label{eq:Dpsi}
D^{\psi\psi^*}(q,\omega_{\nu^*}) \equiv \frac{2\pi/N}{\omega_{\nu^*+1}-\omega_{\nu^*}},
\end{equation}
where $\nu^*=m$ marks the antipsinon quantum number in the $\psi\psi^*$ continuum and
picks the dynamically relevant branch from the set of $2m$-psinon states. The
psinon quantum number $\nu$ is adjusted to keep the wave number $q$ of the $\psi\psi^*$
state fixed. This choice of labels produces an ordered sequence of levels.
Starting at $\omega=0$, the graph of $D^{\psi\psi^*}(\pi/2,\omega_{\nu^*})$ rises from a nonzero
value very slowly up to near the upper band edge, where it bends into a
square-root divergence as shown in Fig.~\ref{fig:8}(a). The divergence is
produced by a maximum of the sequence $\omega_{\nu^*}$ at the fold of the
$\psi\psi^*$-continuum.\cite{note5}

%%%%%%%%%%%%%%%%%%%%%%%%%%%%%%%BEGIN-FIGURE%%%%
%%%%%%%%%%%%%%%%%%%%%%%%%%%%%%%%END-FIGURE%%%%%

In Fig.~\ref{fig:8}(b) we show finite-$N$ data at $q=\pi/2$ for the scaled transition rates
\begin{equation}\label{eq:Mpsi}
M_{zz}^{\psi\psi^*}(q,\omega_{\nu^*}) \equiv N|\langle G|S_{q}^{z}|\nu^*\rangle|^2.
\end{equation}
These data compellingly suggest the existence of a smooth function
$M_{zz}^{\psi\psi^*}(\pi/2,\omega)$ for the $\psi\psi^*$ transition rates in the limit $N\to\infty$, which
further highlights the physical significance of the psinon and the antipsinon as
relevant quasi-particles in this situation.  The function $M_{zz}^{\psi\psi^*}(\pi/2,\omega)$
is monotonically decreasing with a divergence at $\omega=0$ and a cusp singularity at
the upper band edge $\omega_U\simeq1.679J$.

The product of the transition rate function and the (interpolated) density of
states is shown in Fig.~\ref{fig:8}(c).\cite{note8} The curve fitted through the
data points represents the $\psi\psi^*$ lineshape at $q=\pi/2$ in $S_{zz}(q,\omega)$. Its
most distinctive feature is the double peak due to apparent divergences at both
band edges.

The divergence at $\omega=0$, which is caused by the matrix elements, is a power law,
$~\omega^{-\alpha}$, with an exponent that is exactly known from field theoretic studies
of the Heisenberg model.\cite{Hald80,BIR87,FGM+96} For the situation at hand,
the value is $\alpha=0.4688\ldots$.  The divergence at $\omega_U$ is caused by the diverging
density of states but is weakened if the cusp singularity of
$M_{zz}^{\psi\psi^*}(\pi/2,\omega)$ starts from zero at $\omega=\omega_U$. The expectation is a
power-law singularity, $~(\omega_U-\omega)^{-\beta}$ with an exponent $0\leq \beta\leq\frac{1}{2}$.

It is interesting to compare the $\psi\psi^*$ transition rate function
$M_{zz}^{\psi\psi^*}(\pi/2,\omega)$ at $M_z=N/4$ inferred from the Bethe ansatz with the
2-spinon transition rate function $M_{zz}^{(2)}(\pi,\omega)$ at $M_z=0$ calculated
via algebraic analysis.\cite{KMB+97} The shape of both functions is similar, but
there are some differences: $M_{zz}^{(2)}(\pi,\omega)$ has a stronger power-law
divergence at $\omega=0$ and it approaches zero more rapidly at the upper band edge.
As a result it produces a monotonically decreasing spectral-weight distribution
$S_{zz}^{(2)}(\pi,\omega)$ (see Fig.~\ref{fig:1}) notwithstanding the fact that the
2-spinon density of states is also a monotonically increasing function
terminating in a square-root divergence.

At $q=\pi/4$ the integrated intensity $S_{zz}(q)$ is only a third of what it was
at $q=\pi/2$, but spread over a narrower range of frequencies (see
Fig.~\ref{fig:6}). The bandwidth has shrunk to less than a third of the value it
had at $q=\pi/2$. The relative $\psi\psi^*$ contribution to the intensity is even larger
than at $q=\pi/2$, almost 100\%.  In this application, the method of analysis is
stretched more closely to its limits because $q=\pi/4$ exists in fewer manageable
system sizes. However, the data still make reliable lineshape predictions possible.

\begin{figure}[tb]
  \centerline{\epsfig{file=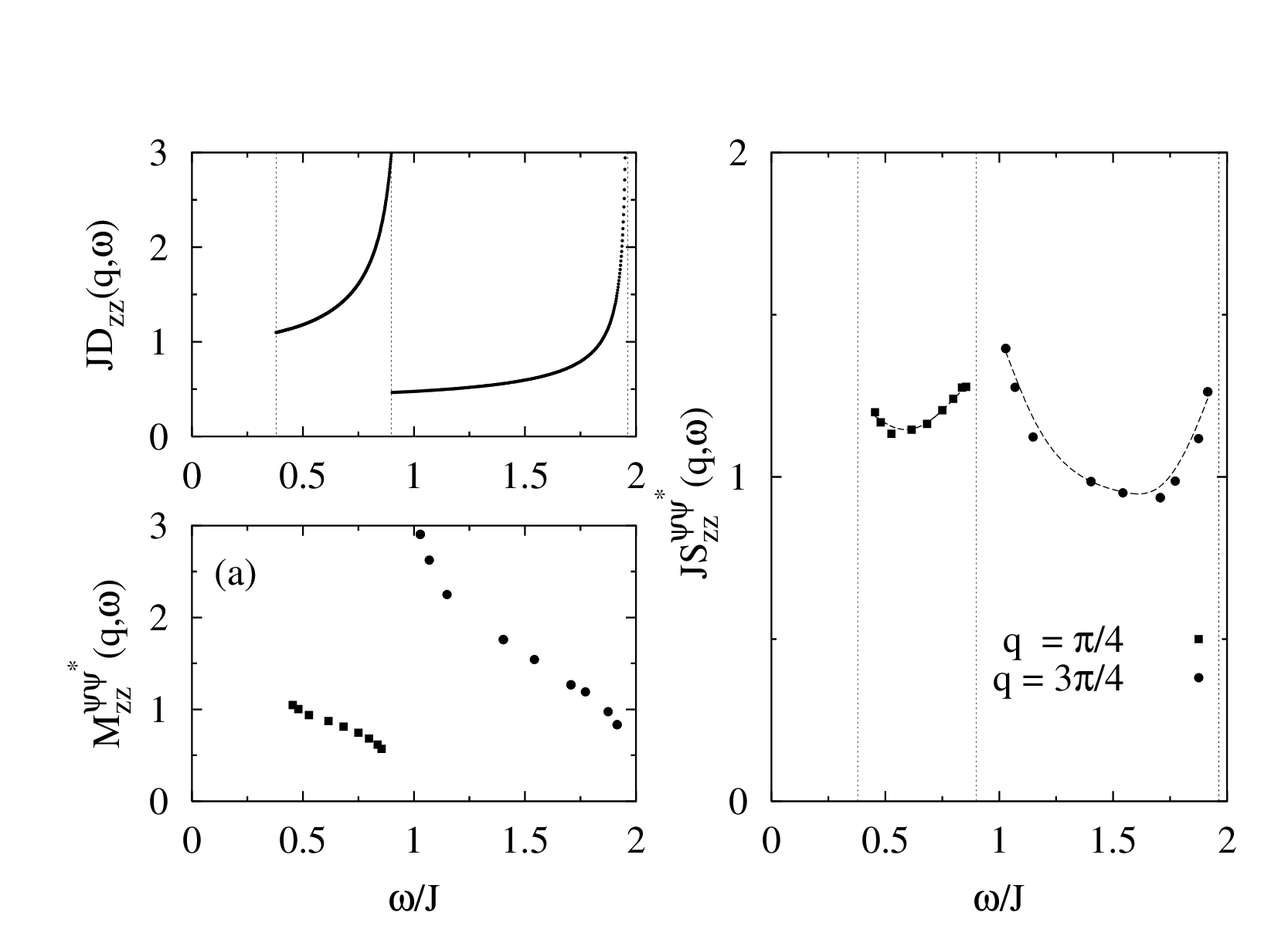,width=6.2cm,angle=-90}}

\caption{(a) Density of $\psi\psi^*$ states at $q=\pi/4,3\pi/4$ evaluated via
  \eqref{eq:Dpsi} from Bethe ansatz data for $N=2048$. (b) Transition rates
  \eqref{eq:Mpsi} between the psinon vacuum and the $\psi\psi^*$ states at
  $q=\pi/4,3\pi/4$ for $N=16,24,32$. (c) Lineshape at $q=\pi/4,3\pi/4$ of the $\psi\psi^*$
  contribution to $S_{zz}(q,\omega)$. All results are for $M_z=N/4$.}
\label{fig:9}
\end{figure}

The density of states $D_{zz}^{\psi\psi^*}(\pi/4,\omega)$, plotted in Fig.~\ref{fig:9}(a),
rises discontinuously from zero to a finite value at the spectral threshold,
$\Delta E \simeq 0.379J$. From there it increases gradually with gradually increasing slope
and ends in a cusp singularity at the upper band edge.\cite{note6}  The
finite-$N$ data for the scaled transition rates shown in Fig.~\ref{fig:9}(b)
again suggest a smooth $\omega$-dependence in the form of a monotonically decreasing
curve with enhanced steepness near both band edges.  However, the countertrend
of the density of states at the upper band edge is of sufficient strength to
produce a second maximum in the line shape again.

%%%%%%%%%%%%%%%%%%%%%%%%%%%%%%%BEGIN-FIGURE%%%%
%%%%%%%%%%%%%%%%%%%%%%%%%%%%%%%%END-FIGURE%%%%%

Also shown in Fig.~\ref{fig:9} are the corresponding data for the $\psi\psi^*$ density
of states, transition rates, and lineshape at $q=3\pi/4$.  Here the relative
spectral weight carried by the $\psi\psi^*$ excitations is only 83\% of the value at
$q=\pi/2$, but that fraction is concentrated over a frequency band that has shrunk
to 65\% of the width at $q=\pi/2$, while the absolute intensity remains fairly high
(87\% of the value at $q=\pi/2$).  Both quantities, which determine the $\psi\psi^*$
lineshape, exhibit similar frequency dependences as we have already observed for
the other two fixed-$q$ scans.  The density of states is divergent again at the
upper boundary. The energy gap is now much larger, $\Delta E\simeq 0.899J$. The fact that
the lower continuum boundary at $q=3\pi/4$ coincides with the upper continuum
boundary at $q=\pi/4$ is a consequence of the quasi-particle dispersions as
discussed previously.

%%%%%%%%%%%%%%%%%%%%%%%%%%%%%%%%%%%%%%%%%%%%%%%
%
\section{Conclusion}\label{sec:VII}
%  
%%%%%%%%%%%%%%%%%%%%%%%%%%%%%%%%%%%%%%%%%%%%%%%
The spectrum of the completely integrable 1D $s=\frac{1}{2}$ Heisenberg
antiferromagnet \eqref{eq:Hh} can be generated in more than one way from
multiple excitations of quasi-particles. The external magnetic field controls
the nature of the ground state. In strong fields, it becomes the vacuum of
magnons and in zero field the vacuum of spinons. The dynamically relevant
collective excitations of specific quantum fluctuations in the two cases are
then naturally described as composites of quasi-particles from the respective
species and are likely to involve only a small number of quasi-particles.

In intermediate magnetic fields, neither the magnons nor the spinons provide a
useful interpretation of dynamically relevant collective excitations for the
same fluctuation operators. The ground state itself contains a macroscopic
number of quasi-particles from one or the other of the two species. However,
when it is reconfigured as the physical vacuum for psinons and antipsinons, then
it turns out that the spin fluctuation operator $S_q^z$ induces predominantly
transitions to $\psi\psi^*$ states, which contain just one particle from each kind.

Like the magnon and the spinon, the psinon and the antipsinon are interacting
quasi-particles in the Heisenberg model \eqref{eq:Hh}. In the $\psi\psi^*$ scattering
states, the interaction energy of the psinon and the antipsinon is of order
O($N^{-1}$) whereas the interaction energy among magnons or spinons is of order
O(1). Hence, for $N\to\infty$, the $\psi\psi^*$ states join up in $(q,\omega)$-space to form a two-parameter
continuum whose spectral boundaries and density of states are fully determined by
the energy-momentum relations of the psinon and the antipsinon. Moreover, the scaled
$\psi\psi^*$ transition rates converge for $N\to\infty$ toward a smooth function of $q$ and $\omega$.

We have exploited these asymptotic quasi-particle properties to extract
lineshape information for the dynamic structure factor $S_{zz}(q,\omega)$, which
probes the spin fluctuations parallel to the applied magnetic field. The same
quasi-particles will also play a dominant role in the spin fluctuations
perpendicular to the field, but here different combinations of them make up the
composition of the dynamically relevant collective excitations. In the dynamic
spin structure factor $S_{-+}(q,\omega)$, for example, the spectral weight is almost
completely carried by 2-psinon excitations.\cite{note3}

In all likelihood, the psinon quasi-particles will also be useful for the
analysis of thermal spin fluctuations in this model system. The peculiar
spectral weight distributions found in recent complete diagonalization
studies\cite{FLS97,FL98} of $S_{zz}(q,\omega)$ at $h=0$ and $T>0$, for example,
indicate the presence of stringent selection rules between collective states
coupled by the spin fluctuation operator $S_q^z$. In zero field, psinon vacua
are densely spread across the entire energy range of the model. Each psinon
vacuum can be used as the reference state of a $2m$-psinon expansion
\eqref{eq:I2msp}. If there are general selection rules related to psinon
quasi-particles among transition rates $|\langle\lambda'|S_q^z|\lambda|^2$ within a given class
$K_r$ of Bethe ansatz solutions, they will have a strong impact on the spectral
weight distribution in $S_{zz}(q,\omega)$ at all temperatures.

%%%%%%%%%%%%%%%%%%%%%%%%%%%%%%%%%%%%%%%%%%%%%%%
%
\acknowledgments
%
%%%%%%%%%%%%%%%%%%%%%%%%%%%%%%%%%%%%%%%%%%%%%%%
We thank Andreas Kl{\"u}mper, Klaus Fabricius, and Alexander Meyerovich for
interesting and useful discussions. Financial support from the URI Research
Office (for G.M.) and from the DFG Schwerpunkt \textit{Kollektive
  Quantenzust{\"a}nde in elektronischen 1D {\"U}bergangsmetallverbindungen} (for M.K.)
is gratefully acknowledged.

%%%%%%%%%%%%%%%%%%%%%%%%%%%%%%%%%%%%%%%%%%%%%%%
%\begin{thebibliography}{100}
%\bibliography{../references,notes}                 %gerhard
%\bibliography{/home/karbach/REFERENCES/references,notes}    %michael

\begin{thebibliography}{10}
\bibitem{note2}
\bibinfo{note}{In some collective excitations, the quasi-particles form bound
  states, and the interaction energy remains significant for $N\to\infty$ even
  if only few quasi-particles are present (bound magnons are discussed in
  Ref.~\onlinecite{KM97}.)}.
\bibitem{note7}
\bibinfo{note}{In the Haldane-Shastry model, where the spinons are
  noninteracting quasi-particles, the need for introducing new quasi-particles
  is less urgent (see Ref.~\onlinecite{TH94})}.
\bibitem{note4}
\bibinfo{note}{This kind of factorization was also used in
  Ref.\onlinecite{KMB+97} for the calculation of the exact 2-spinon part of
  $S_{zz}(q,\omega)$ at $M_z=0$ via algebraic analysis and in a Lanczos study
  of $S_{zz}(q,\omega)$ for the Heisenberg and Haldane-Shastry models
  (Ref.~\onlinecite{LR96}).}
\bibitem{note5}
\bibinfo{note}{Not included in Fig.~\ref{fig:8}(b) is the contribution to
  $D^{\psi\psi^*}(\pi/2,\omega_{\nu^*})$ of the $\psi\psi^*$ states in the
  narrow strip of the continuum that is folded over, because no fold exists in
  systems with $N\leq 32$, for which we have transition rate data available.}
\bibitem{note8}
\bibinfo{note}{It incorporates only the transition rate data from states whose
  excitation energies fall between the $\psi\psi^*$ continuum boundaries. That
  excludes a few data points from Fig.~\ref{fig:8}(b) near the upper band edge
  because of the residual quasi-particle interaction.}
\bibitem{note6}
\bibinfo{note}{At $q=\pi/4$ the $\psi\psi^*$ continuum is not folded at the
  upper boundary.}
\end{thebibliography}
%\bibliographystyle{apsrev}
%\end{thebibliography}

%%%%%%%%%%%%%%%%%%%%%%%%%%%%%%%%%%%%%%%%%%%%%%%
\end{document}